\documentclass{jetpl}
\twocolumn
\usepackage{graphicx}
\usepackage{cmap}
\lat
\usepackage{color}

\title{Nonexponential photoluminescence dynamics in an inhomogeneous ensemble of excitons in WSe$_2$ monolayers}

\rtitle{Nonexponential photoluminescence \ldots}

\sodtitle{Nonexponential photoluminescence dynamics in an inhomogeneous ensemble of excitons in WSe$_2$ monolayers}

\author{M.\,A.\,Akmaev$^{+}$\/\thanks[1]{e-mail: akmaevma@lebedev.ru}, M.\,V.\,Kochiev$^{+}$, A.\,I.\,Duleba$^{+*}$, M.\,V.\,Pugachev$^{+*}$, A.\,Yu.\,Kuntsevich$^{+}$, V.\,V.\,Belykh$^{+}$\/\thanks[2]{e-mail: belykh@lebedev.ru}}

\rauthor{M.\,A.\,Akmaev, M.\,V.\,Kochiev, A.\,I.\,Duleba et al.}

\sodauthor{Akmaev, Kochiev, Duleba}

\address{$^+$Lebedev Physical Institute, Russian Academy of Sciences, Moscow, 119991 Russia\\~\\
$^*$Moscow Institute of Physics and Technology, Dolgoprudnyi, Moscow region, 141700 Russia}

\abstract{The spectral and spatiotemporal dynamics of photoluminescence in monolayers of transition metal dichalcogenide WSe$_2$ obtained by mechanical exfoliation on a Si/SiO$_2$ substrate is studied over a wide range of temperatures and excitation powers. It is shown that the dynamics is nonexponential and, for times $t$ exceeding $\sim$50~ps after the excitation pulse, is described by a dependence of the form $1/(t+t_0)$. Photoluminescence decay is accelerated with a decrease in temperature, as well as with a decrease in the energy of emitting states. It is shown that the observed dynamics cannot be described by a bimolecular recombination process, such as exciton--exciton annihilation. A model that describes the nonexponential photoluminescence dynamics by taking into account the spread of radiative recombination times of localized exciton states in a random potential gives good agreement with experimental data.}

\begin{document}

\maketitle
\paragraph{Introduction.}
Atomically thin layers of transition metal dichalcogenides (TMDCs) are a new class of semiconductor materials that has been actively investigated over the past years \cite{Geim2013,Wang2012,Liu2016,Mak2016,Manzeli2017,Wang2018,Chernozatonskii2018,Vdovin2018,Pekh2020}. These materials include substances with an \textit{MX}$_2$ composition, where \textit{M} is a transition metal (\textit{M} = W, Mo), and \textit{X} is a chalcogen (\textit{X} = S, Se, Te). TMDCs acquire unique properties upon a transition from a bulk crystal to a monolayer. While multilayer TMDCs are indirect-gap semiconductors, monolayer TMDCs feature a direct optical transition at the band gap. The exciton binding energy in TMDC monolayers is about 200-500 meV, so that excitons form the ground energy state at room temperature (see \cite{Wang2018} for a review). The unique properties of these compounds, as well as the possibility of creating heterostructures by combining monolayers of different materials \cite{Geim2013,Liu2016}, make them promising candidates for various uses in optoelectronics \cite{Wang2012,Liu2016,Mak2016}. 

TMDC monolayers possess extreme two-dimensionality, which, in combination with a high contrast in the dielectric constants of the monolayer and its environment, leads to the modification of the carrier-carrier interaction potential \cite{Chernikov2014,Robert2018,Prazdnichnykh2020}. Also TMDCs have an unusual band structure, which is characterized by spin-valley coupling and strong spin-orbit interaction \cite{Glazov2015}. For this reason, it is not always possible to directly extrapolate experience gained in the studies of traditional semiconductor systems with quantum wells, and there are still many open questions regarding the properties of TMDC monolayers. For example, mechanisms responsible for the dynamics of photoluminescence (PL) and, in particular, for exciton recombination are not yet clear. Nonexponential dynamics in the decay of PL and photoinduced reflection or transmission is observed in mechanically exfoliated monolayers directly deposited on a substrate \cite{Sun2014,Kumar2014,Mouri2014,Shin2014,Yu2016,Plechinger2017,Lee2018}. In the above publications, this behavior was attributed to bimolecular recombination processes, in particular, exciton-exciton annihilation, which manifests itself at high pump levels at the initial stage of the decay dynamics. At the same time, exponential dynamics is observed in monolayers encapsulated between hexagonal boron nitride (h-BN) layers  \cite{Robert2017,Cadiz2018,Hoshi2017}, which was attributed to the suppression of exciton-exciton annihilation \cite{Hoshi2017}. 

In this work, we study the PL dynamics in WSe$_2$ monolayers placed directly on a Si/SiO$_2$ substrate. On a long time scale, we observe nonexponential decay of the PL that is well described by an inverse proportionality relation $\sim1/(t+t_0)$ and accelerates as the temperature decreases. We show that, contrary to common belief, the observed kinetics is unrelated to exciton-exciton annihilation and can be explained in terms of emission from an ensemble of exciton states with an inhomogeneous distribution of PL decay times. The decay times of exciton states feature positive correlation (i.e., increase) with their energy, which is characteristic of an ensemble of localized excitons with a spread in the localization length.

\begin{figure*}
\begin{center}
\includegraphics[width=0.8\linewidth]{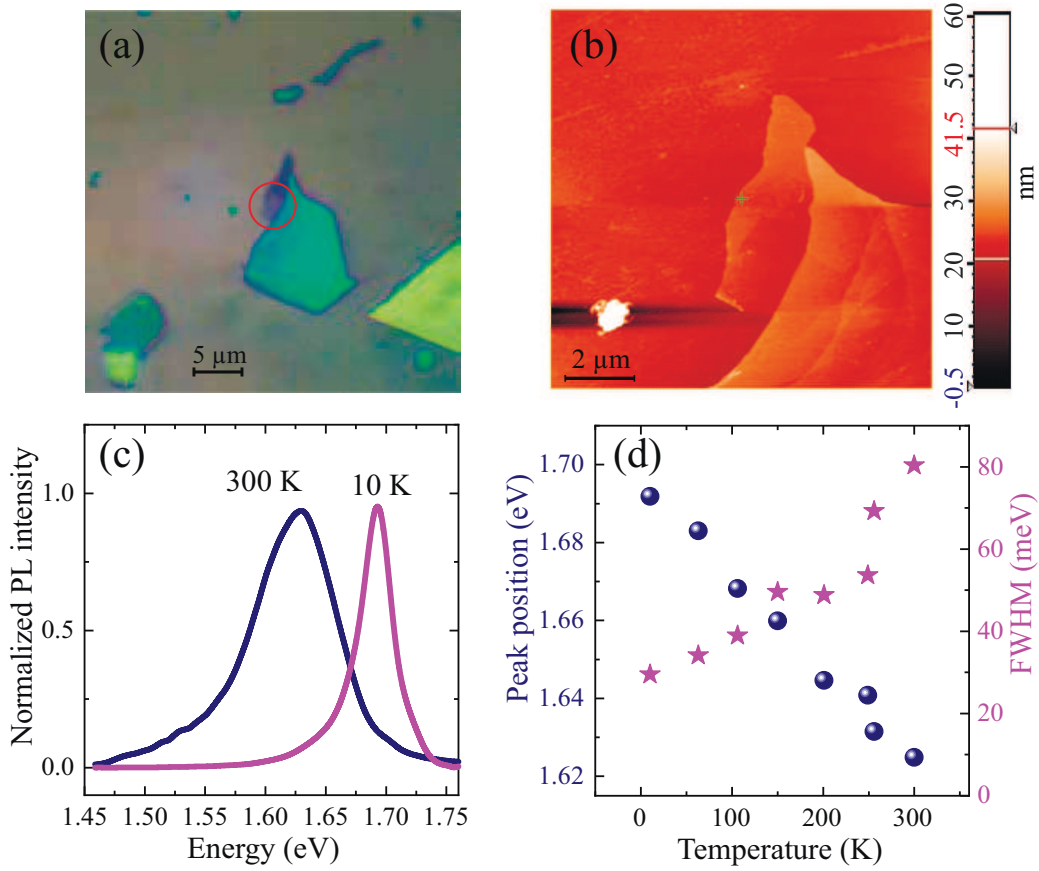}
\caption{Fig. 1. (a) Optical image of a WSe$_2$ monolayer. (b) Image of a WSe$_2$ monolayer obtained with an atomic force microscope. (c) Photoluminescence spectra of a WSe$_2$ monolayer at temperatures of 10 and 300 K. Luminescence has excitonic origin. The curves are normalized to their peak values. (d) Temperature dependences of the peak position and width of the exciton line (circles and stars, respectively).}
\end{center}
\label{fig:Stat}
\end{figure*}

\paragraph{Sample and experimental techniques.}
WSe$_2$ crystals were exfoliated using adhesive tape and transferred to a Si substrate coated with a 285-nm-thick SiO$_2$ layer. Preliminarily, binary marks were applied to the substrate using optical lithography and chromium deposition, accelerating the search for flakes in the future. The initial search for monolayers was carried out using an optical microscope by color. The surface topography of the candidates in monolayer flakes was examined using an NT-MDT Solver 47 atomic force microscope (AFM) in the semi-contact mode. Then the sample was transferred to the setup for studying steady-state PL spectra. Figures 1a and 1b show the optical image and AFM topography of the selected flake, respectively. The lateral size of this flake is about 3 $\mu$m. According to AFM data, the step height is about 1 nm. The PL spectra confirm that this flake is a monolayer one.

The main methods for studying the monolayer WSe$_2$ flakes were steady-state and time-resolved PL in the temperature range of 10-300 K. The sample was placed in vacuum on the cold finger of a helium gas-flow cryostat. To achieve a micrometer spatial resolution, pump laser radiation was focused onto the sample using a microobjective lens, which was also used for collecting PL. In the steady-state PL measurements, the sample was excited by a CW semiconductor laser with a wavelength of 457 nm. The PL spectra were recorded with a resolution of 0.5 meV using a spectrometer with a silicon CCD matrix cooled with liquid nitrogen. In the time-resolved PL measurements, the sample was pumped at a wavelength of 400 nm by the second harmonic of radiation from a pulsed Ti:sapphire laser with a pulse duration of 2 ps. Laser radiation was focused onto the sample in a spot with a diameter of 2 $\mu$m. PL was recorded by a Hamamatsu streak camera combined with a spectrometer. For spectrally-resolved and spatially-resolved measurements, the spectrometer grating was set to the first or zero diffraction order, respectively. The time and spectral resolution in these experiments were up to 5 ps and 1.5 meV, respectively.

\paragraph{Results and discussion.}

Figure 1c shows the PL spectra of the WSe$_2$ monolayer obtained at temperatures of 300 and 10 K. The position and shape of the spectral lines coincide with the literature data for exciton PL \cite{Mouri2014,Zhao2013,Zeng2013,Tonndorf2013,Yan2014,Huang2016,Jadczak2017,Arora2015}. We note that the low-temperature PL spectra reported in \cite{Yan2014,Huang2016,Jadczak2017,Arora2015} exhibit additional low-energy lines that are not observed in our case and are, apparently, associated with defects in the prepared layers.
We investigate WSe$_2$ monolayers placed directly on a Si/SiO$_2$ substrate without h-BN encapsulation. As a result, the spectrum is inhomogeneously broadened even at low temperatures, which does not allow separating the contributions of neutral and charged exciton to the PL \cite{Cadiz2017}. The presence of trion PL is indicated by the asymmetry of the spectral line. Figure 1d shows the temperature dependence of the position and width of the PL line. As the temperature increases, we observe red shift and line broadening associated with a decrease in the band gap and with thermal broadening, respectively. 

\begin{figure*}
\begin{center}
\includegraphics[width=0.8\linewidth]{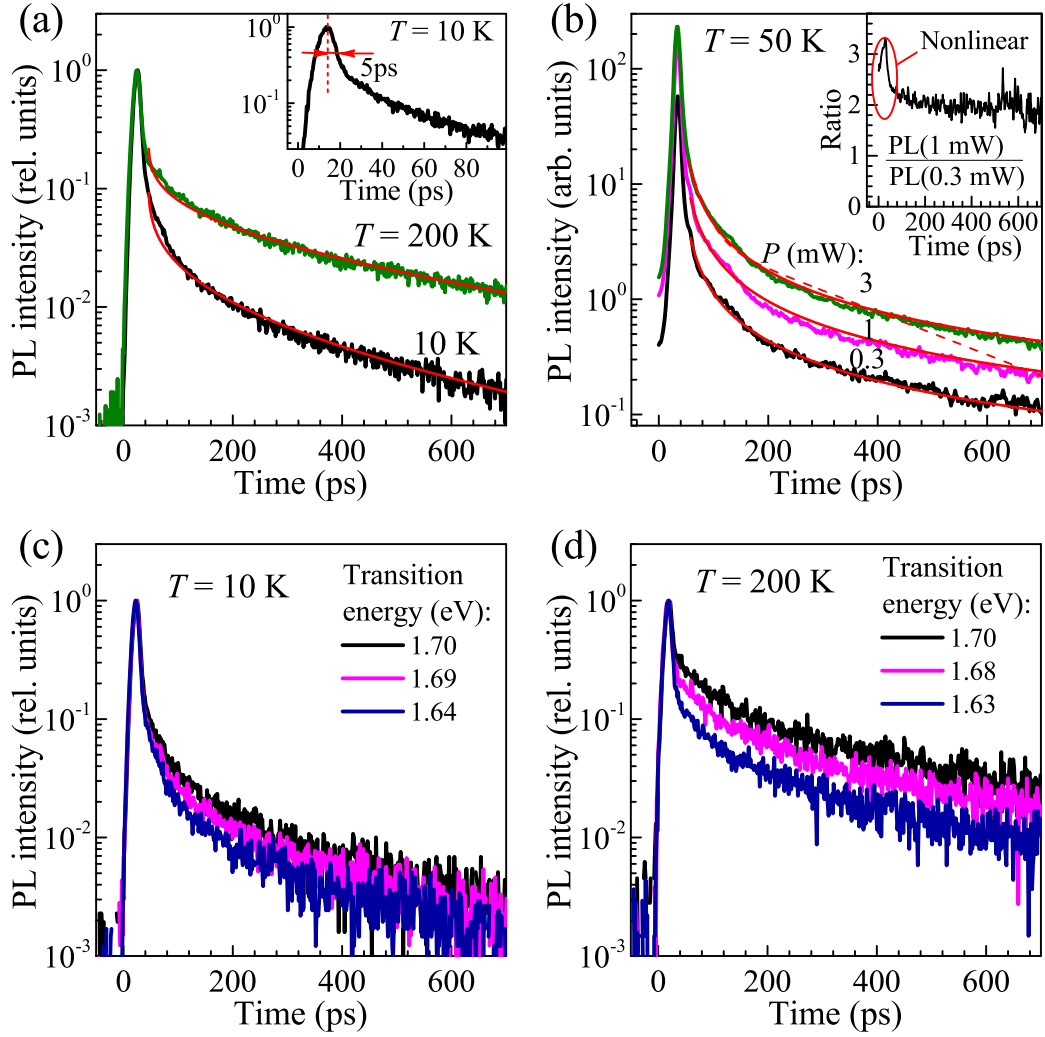}
\caption{Fig. 2. Photoluminescence dynamics in WSe$_2$ monolayers. (a) Photoluminescence dynamics at different temperatures and a pump power of \textit{P} = 1 mW. Red curves show the results of calculations according to Eq. (5). The inset shows the initial stage of the dynamics. (b) Dynamics at different pump powers and a temperature of 50 K. The inset shows the ratio of the intensities corresponding to pump powers of 1 and 0.3 mW. (c) and (d) Photoluminescence dynamics for different energies of the optical transition. Curves in panels (a), (c), and (d) are normalized to their peak values.}
\end{center}
\label{fig:Dyn}
\end{figure*}
Figure 2a shows the dynamics of emission from a WSe$_2$ monolayer at different temperatures upon pulsed laser excitation at a wavelength of 400 nm. The dynamics is nonexponential with a pronounced fast stage after which the decay rate decreases. The initial part of the dynamics is shown in more detail in the inset to Fig. 2a. The fastest component of the dynamics is characterized by a decay time of less than 5 ps. The contribution of the fast component increases (the decay kinetics accelerates) as the temperature decreases. Also, the contribution of the fast component increases with a decrease in the optical transition energy where the PL is detected; this is true for both low (Figs. 2c) and high (Figs. 2d) temperatures.

The emission dynamics for different excitation powers is shown in Fig. 2b. Qualitative changes in the dynamics are difficult to distinguish: it remains nonexponential for both high and low excitation powers. 

\subparagraph{Checking the bimolecular recombination hypothesis.}

The nonexponential PL dynamics at times $t\gtrsim 50$~ps is well described by the dependence $I(t) \sim 1/(t + t_0)$, where $t_0$ is a constant. Fits of kinetic curves by functions of this form are shown in Fig. 2b by red solid lines. For comparison, the dashed line shows the best fit with a biexponential function over the same time interval for a pump power of $P = 3$ mW; however, this curve deviates significantly from the experimental data. The time dependence of the PL intensity of the form $I(t) \sim 1/(t + t_0)$ was observed in many studies for monolayers of WSe$_2$ \cite{Mouri2014,Plechinger2017,Wang2019}, WS$_2$ \cite{Yu2016,Plechinger2017,Lee2018}, MoSe$_2$ \cite{Kumar2014,Shin2014} and MoS$_2$ \cite{Sun2014,Yu2016,Plechinger2017}. This dependence was explained by the impact of a bimolecular process requiring the participation of two excitons in recombination. One such process is exciton-exciton annihilation (Auger recombination), whereby one of the excitons recombines nonradiatively transferring energy to the second exciton, which can dissociate. The dynamics of the exciton concentration $n$ upon bimolecular recombination is described by the equation
\begin{equation}
\frac{dn}{dt} = -C n^2 - \frac{n}{\tau}.
\end{equation}
Here, $C$ is is the constant that determines the rate of bimolecular recombination, and the second (linear) term describes the radiative recombination of excitons with a time constant $\tau$ and determines the PL intensity $I(t) = n(t) / \tau$. The solution to this equation is
\begin{equation}
n(t) = \left[(C\tau+1/n_0)\exp(t/\tau) - C\tau\right]^{-1},
\end{equation}
where $n_0$ is the initial concentration of excitons. When the contribution of the bimolecular process is dominant, which is the case for $t \ll \tau$ and $C n_0 \tau \gg 1$, the exciton concentration depends on time as $n(t) \approx n_0 / [1 + C n_0 t]$, whereas in the limiting case of $t \gg \tau$ the exciton concentration decays exponentially: $n(t) = (C\tau+1/n_0)^{-1}\exp(-t/\tau)$. Note that we observe no transition to exponential decay even for times as long as $t=600$. Thus, if the bimolecular process is really dominant, condition $t \ll \tau$ should be valid. In this case, the ratio of PL intensities for two different excitation powers and, respectively, different initial exciton concentrations $n_0$ and $\tilde{n}_0$ should decrease with time from the value $\tilde{n}_0/n_0$ to 1: $\tilde{I}(t)/I(t) =1 + (\tilde{n}_0/n_0-1)/(C\tilde{n}_0 t + 1)$. The inset in Fig. 2b shows the ratio of two kinetic curves recorded for pump powers differing by a factor of 3. Evidently, this ratio changes only at the initial stage of the kinetics ($t \lesssim 50$~ps) and is constant at longer times, in the region where the PL intensity varies inversely with time. By fitting the experimental curves with the dependence $I(t)\sim 1/(t+t_0)$, we can determine the expected bimolecular recombination coefficient $C = -(dn/dt)/n^2$. Its dependence on the excitation power is shown in Fig. 3a. The coefficient $C$ is plotted in arbitrary units; only its relative change in various experiments has physical meaning. An increase in the excitation power is equivalent to an increase in the initial concentration and should not be accompanied by a significant change in the coefficient $C$. Meanwhile, our measurements show that $C \sim 1/P$. Figure 3b shows the dependence of $C$ on the optical transition energy at two different temperatures. Contradictions appear in this case as well: contrary to our observations, an increase in energy or temperature should lead to the delocalization of excitons and an increase in the efficiency of the bimolecular process, i.e., an increase in $C$ \cite{Robert2016}. Finally, the exciton spatial distribution should change considerably in the case of bimolecular recombination. Areas with higher initial exciton concentration should be emptied more rapidly, which should lead to the efficient broadening of the spatial distribution of excitons and broadening of the PL spot. The diffusion of excitons should only enhance this broadening. The measured and calculated spatial distributions of the PL normalized to the peak value are shown in Fig. 3c for different instants in time. Experimentally, we observe no significant increase in the width of the PL spot, while the calculation predicts considerable broadening, which, however, should be limited by the size of the monolayer flake. Therefore, in our case, the dynamics of the PL intensity decay at times $t \gtrsim 50$~ps, and, in particular, the time dependence $I(t)\sim 1/(t+t_0)$, cannot be explained in terms of a bimolecular process and, apparently, has a different nature.

\begin{figure*}
\begin{center}
\includegraphics[width=1\linewidth]{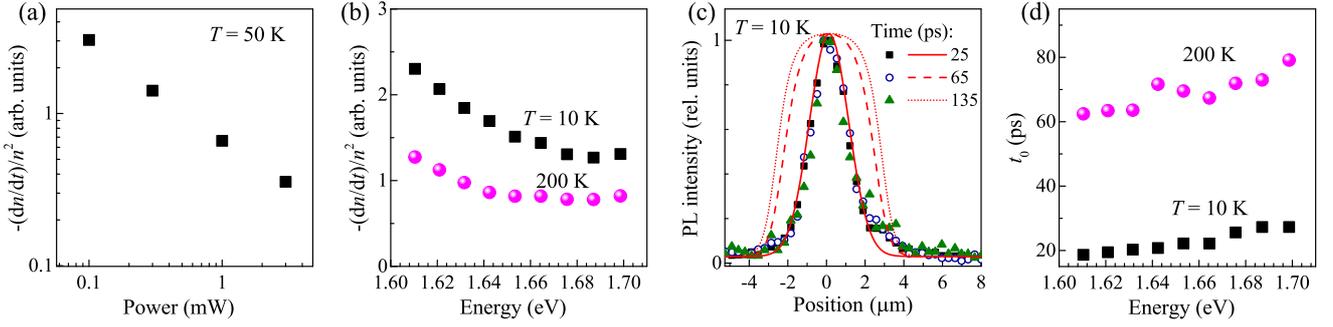}
\caption{Fig. 3. (a)-(c) Check of the bimolecular recombination hypothesis. (a) Dependence of the expected bimolecular recombination coefficient $C = -(dn/dt)/n^2$ at a temperature of 50 K on the pump power. (b) Dependence of the expected bimolecular recombination coefficient on the emission energy at two different temperatures. (c) Photoluminescence spot profiles at different times after the excitation pulse at $T = 10$ K normalized to their peak values. Symbols and lines show the experimental and calculated results. (d) Dependence of the characteristic time $t_0$ of the photoluminescence dynamics on the emission energy at different temperatures.}
\end{center}
\label{fig:Bi}
\end{figure*}

\subparagraph{Linear model of the nonexponential dynamics}
The fact that the character of the PL dynamics at $t \gtrsim 50$~ps is independent of the excitation power suggests that the dynamics of the exciton concentration is described by linear equations. In this case, the nonexponential character of the intensity decay with time is caused by the fact that we observe the intensity of emission from an inhomogeneous ensemble of states where each state exhibits exponential decay but the decay time $\tau$ is different for different states. 
Then,
\begin{equation}
I(t) = \int_0^\infty \frac{n_0(\tau)}{\tau}\exp(-t/\tau)d\tau, 
\end{equation} 
where $n_0(\tau)d\tau$ is the concentration of excitons at time $t = 0$ in states characterized by decay time $\tau$ in the interval $d\tau$. In particular, to obtain the dependence $I(t) \propto 1/(t+t_0)$, which is close to the experimental one, the function $n_0(\tau)$ should have the form $n_0(\tau) \propto n_0\exp(-t_0/\tau)/\tau$, where the time $t_0$ corresponds to the maximum of the distribution $n_0(\tau)$ and may be considered the characteristic time of the PL dynamics. The dependence of this time on the energy of the emitting state at different temperatures is shown in Fig. 3d. This dependence confirms the conclusion that the dynamics slows down with an increase in the energy of the emitting states or temperature.

Nonexponential dynamics associated with the inhomogeneity of recombination times is found in many systems  \cite{Huntley2006,Phillips1996,Jonscher1984,Gilinsky2001,Cardin2013,Morel2003,Brosseau2010,Bartel2004,Krivobok2020}. In the vast majority of them, nonexponential dynamics is associated with the need for electron tunneling toward recombination centers or with donor-acceptor recombination. In this case, recombination should speed up with an increase in temperature, which contradicts our observations. We note that the slowing down of the PL dynamics with increasing temperature was also reported in other studies for WSe$_2$ \cite{Zhang2015,Godde2016}, MoS$_2$ \cite{Moody2016a} and MoSe$_2$ \cite{Godde2016} and, apparently, is general for TMDC monolayers. 

Before we move on to determine the character of the distribution of decay times $\tau$ and find $I(t)$, let us give a few remarks about the nature of emitting states. The exciton ground state in WSe$_2$ monolayers is dark, while the bright state is ~40 meV higher in energy  \cite{Zhang2015,Wang2017}. It was shown in \cite{Wang2017} that the ground state can nevertheless emit at a nonzero angle to the normal. Apparently, this state determines the PL dynamics, at least at low temperatures. A significant role in the PL dynamics in WSe$_2$ monolayers at low temperatures is played by trion states \cite{Godde2016}, which are characterized by short PL decay times. Biexciton states can also contribute to PL in monolayer WSe$_2$. In our case, their contribution to the dynamics at $t\gtrsim 50$~ps is insignificant, since the corresponding intensity would be quadratic as a function of the pump power (see Fig. 2b).

The characteristic recombination time of neutral excitons in TMDCs $\tau_0$ is rather short \cite{Robert2016}. An exciton can emit if its total wave vector lies within the light cone $|k|<\omega/c$ (where $\omega$ is the photon frequency) and, therefore, its kinetic energy is close to zero. In thermal equilibrium, the exciton energies are distributed in the range of $\sim k_\text{B}T$, and the decay time of the total exciton concentration can be estimated as $\tau \sim \tau_0 k_\text{B}T /(\hbar^2\omega^2/2mc^2) \gg \tau_0$ \cite{Belykh2015}, where $m \approx 0.8 m_0$ is the exciton mass \cite{Cadiz2018} and $m_0$ is the free electron mass. This explains the rather long PL dynamics and the increase in the characteristic PL decay time with increasing temperature. In our case of nonencapsulated WSe$_2$ on a Si/SiO$_2$ substrate, there is inhomogeneity associated with fluctuations in the potential landscape where excitons move. This inhomogeneity manifests itself in the broadening of the PL spectrum. Potential fluctuations lead to the localization of excitons and spreading of exciton states in $k$ space on the order of $\delta k^2 \sim 1/L^2$, where $L$ is the characteristic localization length (Fig. 4). Then, the radiative lifetime of a localized state is $\tau \sim \tau_0 c^2 /\omega^2 L^2$. If we assume that the inhomogeneous broadening in the spectrum is caused by the spread in localization lengths $L$, the energy of a localized state (measured from the bottom of the potential well) $E \sim \hbar^2/2mL^2 \sim (\tau/\tau_0) (\hbar\omega)^2/2mc^2$. Therefore, in a system where inhomogeneity is related to a spread in the size of the localizing potential, the characteristic exciton recombination time is
\begin{equation}
\tau / \tau_0 \sim \alpha E, 
\label{eq:tau}
\end{equation}
where $\alpha = 2mc^2/(\hbar\omega)^2 \approx 400$~400 meV$^{-1}$. If, apart from variations in the size of the localizing potential, we also take into account variations in its depth, there appears a spread in the values of $\tau$ corresponding to the same energy in the spectrum, but the overall trend, described by Eq. \eqref{eq:tau}, remains unchanged. For trion states, there is no requirement that the total wave vector be close to zero, because momentum can be transferred to the remaining carrier upon recombination. For this reason, the trion PL decay rate is fairly high \cite{Godde2016} and trions do not contribute significantly to the dynamics at $t \gtrsim 50$~ps. Anyhow, taking into account trion states that have a short PL decay time and are lower in energy fits into the trend of an increase in the decay time with energy, set by Eq.  \eqref{eq:tau}. We note that the increase in the characteristic PL decay time with an increase in the energy of the emitting states is confirmed by experimental data (Figs. 2c, 2d).
\begin{figure}
\begin{center}
\includegraphics[width=1\columnwidth]{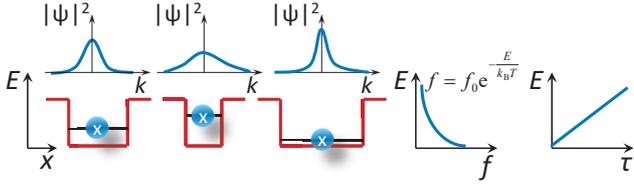}
\caption{Fig. 4. Schematic representation of excitons in localizing potentials of different extensions. For shorter extension of the potential, the exciton energy, the width of the momentum distribution (see upper part of the figure), and the radiative recombination time $\tau$ increase, while the occupancy $f$ of the corresponding states decreases.}
\end{center}
\label{fig:scheme}
\end{figure} 
Let us determine the total PL intensity by adding together the intensities $f_0 \exp(-E/k_\text{B}T) \exp(-t/\tau)/\tau$ of emission from individual states whose occupancy, under the assumption of thermal equilibrium, is described by the Boltzmann distribution $f_0 \exp(-E/k_\text{B}T)$:
\begin{multline}
I(t) = \int_0^\infty \frac{f_0}{\tau} \exp\left(-\frac{E}{k_\text{B}T}\right) \exp\left(-\frac{t}{\tau}\right) g(E) dE = \\
= A\int_0^\infty \exp\left(-\frac{\tau}{\alpha \tau_0 k_\text{B}T} - \frac{t}{\tau}\right) \frac{d\tau}{\tau},
\label{eq:i}
\end{multline}
Here, we use $\tau / \tau_0 = \alpha E$, $f_0$ is a constant that determines the occupancy; $g(E)$ is the density of states, which, for simplicity, can be assumed constant like that of free states in a two-dimensional system; and $A = f_0 g/\alpha\tau_0$.

The red lines in Fig. 2a show the results of calculations according to Eq. ~\eqref{eq:i}. These should be compared with the experimental data for the dynamics of the PL intensity at times $t = 50$; it is reasonable to expect that thermal equilibrium is established in the system and rapid nonlinear processes, including those related to biexcitons and to bimolecular recombination, fade out at this time scale. The calculations use the value of $\tau_0 = 1.3$~ps, which is in reasonable agreement with the available experimental data on the recombination time of excitons with zero wave vector \cite{Robert2016}. Note the good agreement of the calculated curves with experimental data. In particular, the calculations yield nonexponential dynamics that slows down with time and is close to $1/(t+t_0)$ and reproduce changes in the dynamics with temperature. 

A more complete analysis should include taking into account all possible exciton states with different combinations of electron and hole spins, excitons consisting of an electron and a hole occupying different valleys \cite{Wang2017,Li2019,Peng2019}, as well as trion states \cite{Godde2016}. Excitons with different spin or valley structures may not reach thermal equilibrium between them, which further complicates the analysis. Nonetheless, to explain the experimentally observed long-lived PL dynamics at $t \gtrsim 50$~ps, which is nonexponential, slows down with time, is independent of the excitation power, and accelerates with a decrease in temperature, only two conditions have to be met: (i) the observed PL should originate from a set of emitting states characterized by different PL decay times $\tau$ and (ii) there should be positive correlation between the radiative lifetime $\tau$ and the energy $E$ of a given state. Neither a constant density of emitting states nor a linear, or even unambiguous, dependence $\tau(E)$ are required. On a shorter time scale ($t\lesssim 50$~ps), nonlinear processes such as bimolecular recombination can also make significant contribution to the dynamics, as was noted in \cite{Sun2014,Shin2014,Yu2016,Wang2019}.
  
We are grateful to S.N. Nikolaev for providing WSe$_2$ crystals and to M.M. Glazov and M.L. Skorikov for fruitful discussions and valuable advice.
Substrate preparation, flake transfer, and localization of monolayer samples were carried out at the Shared Facility Center of the Lebedev Physical Institute.


\begin{thebibliography}{10}

\bibitem{Geim2013}
A.~K. Geim and I.~V. Grigorieva,
\newblock Nature \textbf{499}, 419 (2013).

\bibitem{Wang2012}
Q.~H. Wang, K. Kalantar-Zadeh, A. Kis, J. N. Coleman and M. S. Strano,
\newblock Nat. Nanotechnol. \textbf{7}, 699 (2012).

\bibitem{Liu2016}
Y.~Liu, N. O. Weiss, X. Duan, H.-C. Cheng, Y. Huang and X. Duan,
\newblock Nat. Rev. Mater. \textbf{1}, 16042 (2016).

\bibitem{Mak2016}
K.~F. Mak and J.~Shan,
\newblock Nat. Photonics \textbf{10}, 216 (2016).

\bibitem{Manzeli2017}
S.~Manzeli, D. Ovchinnikov, D. Pasquier, O. V. Yazyev and A. Kis,
\newblock Nat. Rev. Mater. \textbf{2}, 17033 (2017).

\bibitem{Wang2018}
G.~Wang, A. Chernikov, M. M. Glazov, T. F. Heinz, X. Marie, T. Amand and B. Urbaszek,
\newblock Rev. Mod. Phys. \textbf{90}, 021001 (2018).

\bibitem{Chernozatonskii2018}
L.~A. Chernozatonskii and A.~A. Artyukh,
\newblock Uspekhi Fiz. Nauk \textbf{188}, 3 (2018).

\bibitem{Vdovin2018}
E.~E. Vdovin and Y.~N. Khanin,
\newblock Jetp Lett. \textbf{108}, 641 (2018).

\bibitem{Pekh2020}
P.~L. Pekh, P.~V. Ratnikov and A.~P. Silin,
\newblock Jetp Lett. \textbf{111}, 90 (2020).

\bibitem{Chernikov2014}
A. Chernikov, T. C. Berkelbach, H. M. Hill, A. Rigosi, Y. Li, O. B. Aslan, D. R. Reichman, M. S. Hybertsen and T. F. Heinz,
\newblock Phys. Rev. Lett. \textbf{113}, 076802 (2014).

\bibitem{Robert2018}
C. Robert, M. A. Semina, F. Cadiz, M. Manca, E. Courtade, T. Taniguchi, K. Watanabe, H. Cai, S. Tongay, B. Lassagne, P. Renucci, T. Amand, X. Marie, M. M. Glazov and B. Urbaszek,
\newblock Phys. Rev. Materials 2, 011001(R) (2018).

\bibitem{Prazdnichnykh2020}
A.~I. Prazdnichnykh, M. M. Glazov, L. Ren, C. Robert, B. Urbaszek and X. Marie,
\newblock arXiv:2010.01352 (2020).

\bibitem{Glazov2015}
M.~M. Glazov, E. L. Ivchenko, G. Wang, T. Amand, X. Marie, B. Urbaszek and B. L. Liu,
\newblock Phys. status solidi \textbf{252}, 2349 (2015).

\bibitem{Sun2014}
D.~Sun, Y. Rao, G. A. Reider, G. Chen, Y. You, L. Brézin, A. R. Harutyunyan and T. F. Heinz,
\newblock Nano Lett. \textbf{14}, 5625 (2014).

\bibitem{Kumar2014}
N.~Kumar, Q. Cui, F. Ceballos, D. He, Y. Wang and H. Zhao,
\newblock Phys. Rev. B \textbf{89}, 125427 (2014).

\bibitem{Mouri2014}
S.~Mouri, Y. Miyauchi, M. Toh, W. Zhao, G. Eda and K. Matsuda,
\newblock Phys. Rev. B \textbf{90}, 155449 (2014).

\bibitem{Shin2014}
M.~J. Shin, D.~H. Kim and D.~Lim
\newblock J. Korean Phys. Soc. \textbf{65}, 2077 (2014).

\bibitem{Yu2016}
Y.~Yu, Y. Yu, C. Xu, A. Barrette, K. Gundogdu and L. Cao,
\newblock Phys. Rev. B \textbf{93}, 201111 (2016).

\bibitem{Plechinger2017}
G.~Plechinger, P. Nagler, A. Arora, R. Schmidt, A. Chernikov, J. Lupton, R. Bratschitsch, C. Schüller and T. Korn,
\newblock Phys. status solidi - Rapid Res. Lett. \textbf{11}, 1700131 (2017).

\bibitem{Lee2018}
Y.~Lee, G. Ghimire, S. Roy, Y. Kim, C. Seo, A. K. Sood, J. I. Jang and J. Kim,
\newblock ACS Photonics \textbf{5}, 2904 (2018).

\bibitem{Robert2017}
C.~Robert, T. Amand, F. Cadiz, D. Lagarde, E. Courtade, M. Manca, T. Taniguchi, K. Watanabe, B. Urbaszek and X. Marie,
\newblock Phys. Rev. B \textbf{96}, 155423 (2017).

\bibitem{Cadiz2018}
F.~Cadiz, C.~Robert, E. Courtade, M. Manca, L. Martinelli, T. Taniguchi, K. Watanabe, T. Amand, A. C. H. Rowe, D. Paget, B. Urbaszek and X. Marie,
\newblock Appl. Phys. Lett. \textbf{112}, 152106 (2018).

\bibitem{Hoshi2017}
Y.~Hoshi, T. Kuroda, M. Okada, R. Moriya, S. Masubuchi, K. Watanabe, T. Taniguchi, R. Kitaura and T. Machida,
\newblock Phys. Rev. B \textbf{95}, 241403 (2017).

\bibitem{Zhao2013}
W.~Zhao, Z. Ghorannevis, L. Chu, M. Toh, C. Kloc, P.-H. Tan and G. Eda,
\newblock ACS Nano \textbf{7}, 791 (2013).

\bibitem{Zeng2013}
H.~Zeng, G.-B. Liu, J. Dai, Y. Yan, B. Zhu, R. He, L. Xie, S. Xu, X. Chen, W. Yao and X. Cui,
\newblock Sci. Rep. \textbf{3}, 1608 (2013).

\bibitem{Tonndorf2013}
P.~Tonndorf, R. Schmidt, P. Böttger, X. Zhang, J. Börner, A. Liebig, M. Albrecht, C. Kloc, O. Gordan, D. R. T. Zahn, S. M. de Vasconcellos and R. Bratschitsch,
\newblock Opt. Express \textbf{21}, 4908 (2013).

\bibitem{Yan2014}
T.~Yan, X. Qiao, X. Liu, P. Tan and X. Zhang,
\newblock Appl. Phys. Lett. \textbf{105}, 101901 (2014).

\bibitem{Huang2016}
J.~Huang, T.~B. Hoang and M.~H. Mikkelsen,
\newblock Sci. Rep. \textbf{6}, 22414 (2016).

\bibitem{Jadczak2017}
J.~Jadczak, J. Kutrowska-Girzycka, P. Kapuściński, Y. S. Huang, A. Wójs and L. Bryja,
\newblock Nanotechnology \textbf{28}, 395702 (2017).

\bibitem{Arora2015}
A.~Arora, M. Koperski, K. Nogajewski, J. Marcus, C. Faugeras and M. Potemski,
\newblock Nanoscale \textbf{7}, 10421 (2015).

\bibitem{Cadiz2017}
F.~Cadiz, E. Courtade, C. Robert Collaboration,
\newblock Phys. Rev. X \textbf{7}, 21026 (2017).

\bibitem{Wang2019}
J.~Wang, Y. Guo, Y. Huang, H. Luo, X. Zhou, C. Gu and B. Liu,
\newblock Appl. Phys. Lett. \textbf{115}, 131902 (2019).

\bibitem{Robert2016}
C.~Robert, D. Lagarde, F. Cadiz, G. Wang, B. Lassagne, T. Amand, A. Balocchi, P. Renucci, S. Tongay, B. Urbaszek and X. Marie,
\newblock Phys. Rev. B \textbf{93}, 205423 (2016).

\bibitem{Huntley2006}
D.~J. Huntley,
\newblock J. Phys. Condens. Matter \textbf{18}, 1359 (2006).

\bibitem{Phillips1996}
J.~C. Phillips,
\newblock Reports Prog. Phys. \textbf{59}, 1133 (1996).

\bibitem{Jonscher1984}
A.~K. Jonscher and A.~de~Polignac,
\newblock J. Phys. C Solid State Phys. \textbf{17}, 6493 (1984).

\bibitem{Gilinsky2001}
A.~M. Gilinsky and K.~S. Zhuravlev,
\newblock Appl. Phys. Lett. \textbf{79}, 3455 (2001).

\bibitem{Cardin2013}
V.~Cardin, L. I. Dion-Bertrand, P. Grégoire, H. P. T. Nguyen, M.Sakowicz, Z. Mi, C. Silva and R. Leonelli,
\newblock Nanotechnology \textbf{24}, 045702 (2013).

\bibitem{Morel2003}
A.~Morel, P. Lefebvre, S. Kalliakos, T. Taliercio, T. Bretagnon and B. Gil,
\newblock Phys. Rev. B \textbf{68}, 045331 (2003).

\bibitem{Brosseau2010}
C.-N. Brosseau, M. Perrin, C. Silva and R. Leonelli,
\newblock Phys. Rev. B \textbf{82}, 085305 (2010).

\bibitem{Bartel2004}
T.~Bartel, M. Dworzak, M. Strassburg, A. Hoffmann, A. Strittmatter and D. Bimberg,
\newblock Appl. Phys. Lett. \textbf{85}, 1946 (2004).

\bibitem{Krivobok2020}
V.~S.~Krivobok, A. V. Kolobov, S. E. Dimitrieva, D. F. Aminev, S. I. Chentsov, S. N. Nikolaev, V. P. Martovitskii и E. E. Onishchenko,
\newblock JETP Lett. \textbf{112}, 471 (2020).

\bibitem{Zhang2015}
X.-X. Zhang, Y. You, S. Y. F. Zhao and T. F. Heinz,
\newblock Phys. Rev. Lett. \textbf{115}, 257403 (2015).

\bibitem{Godde2016}
T.~Godde, D. Schmidt, J. Schmutzler, M. Aßmann, J. Debus, F. Withers, E. M. Alexeev, O. Del Pozo-Zamudio, O. V. Skrypka, K. S. Novoselov, M. Bayer and A. I. Tartakovskii,
\newblock Phys. Rev. B \textbf{94}, 165301 (2016).

\bibitem{Moody2016a}
G.~Moody, J.~Schaibley and X.~Xu,
\newblock J. Opt. Soc. Am. B \textbf{33}, C39 (2016).

\bibitem{Wang2017}
G.~Wang, C. Robert, M. M. Glazov, F. Cadiz, E. Courtade, T. Amand, D. Lagarde, T. Taniguchi, K. Watanabe, B. Urbaszek and X. Marie,
\newblock Phys. Rev. Lett. \textbf{119}, 047401 (2017).

\bibitem{Belykh2015}
V.~V. Belykh and M.~V. Kochiev,
\newblock Phys. Rev. B \textbf{92}, 045307 (2015).

\bibitem{Li2019}
Z.~Li, T. Wang, C. Jin Collaboration,
\newblock ACS Nano \textbf{13}, 14107 (2019).

\bibitem{Peng2019}
G.-H. Peng, P.-Y. Lo, W.-H. Li, Y.-C. Huang, Y.-H. Chen, C.-H. Lee, C.-K. Yang and S.-J. Cheng,
\newblock Nano Lett. \textbf{19}, 2299 (2019).

\end{thebibliography}
\end{document}